\documentclass[letterpaper, twoside]{JHEP3}
\usepackage{url}
\usepackage{graphicx}	
\usepackage{amsmath}

\title{Huge entropy production inside black holes}

\author{Colin S Wallace \\
	Dept.\ Astrophysical \& Planetary Sciences, Box 391, \\
	University of Colorado, Boulder CO 80309, U.S.A. \\
	E-mail: \email{Colin.Wallace@colorado.edu}
}
\author{Andrew J S Hamilton \\
	JILA, Box 440, University of Colorado, Boulder CO 80309, U.S.A.\\
	Dept.\ Astrophysical \& Planetary Sciences, Box 391, \\
	University of Colorado, Boulder CO 80309, U.S.A.\\ 
	E-mail: \email{Andrew.Hamilton@colorado.edu}
}
\author{Gavin Polhemus \\
	JILA, Box 440, University of Colorado, Boulder CO 80309, U.S.A. \\
	Poudre High School, 201 Impala Drive, Fort Collins, CO 80521, U.S.A. \\
	E-mail: \email{gavin.polhemus@colorado.edu}
}

\newcommand{\dd}{d}
\newcommand{\geff}{g}

\newcommand{\Msun}{\mbox{$\textrm{M}_\odot$}}
\newcommand{\Mbh}{M_\bullet}
\newcommand{\Mdot}{\dot{M}_\bullet}

\newcommand{\Planck}{\textrm{p}}
\newcommand{\Qbh}{Q_\bullet}

\newcommand{\Scov}{S_\textrm{cov}}
\newcommand{\SBH}{S_\textrm{BH}}
\newcommand{\splat}{\#}

\newcommand{\Vol}{V}

\newcommand{\unit}[1]{\, \textrm{#1}}

\newcommand{\varsfig}{
    \FIGURE{
    \includegraphics[scale=.9]{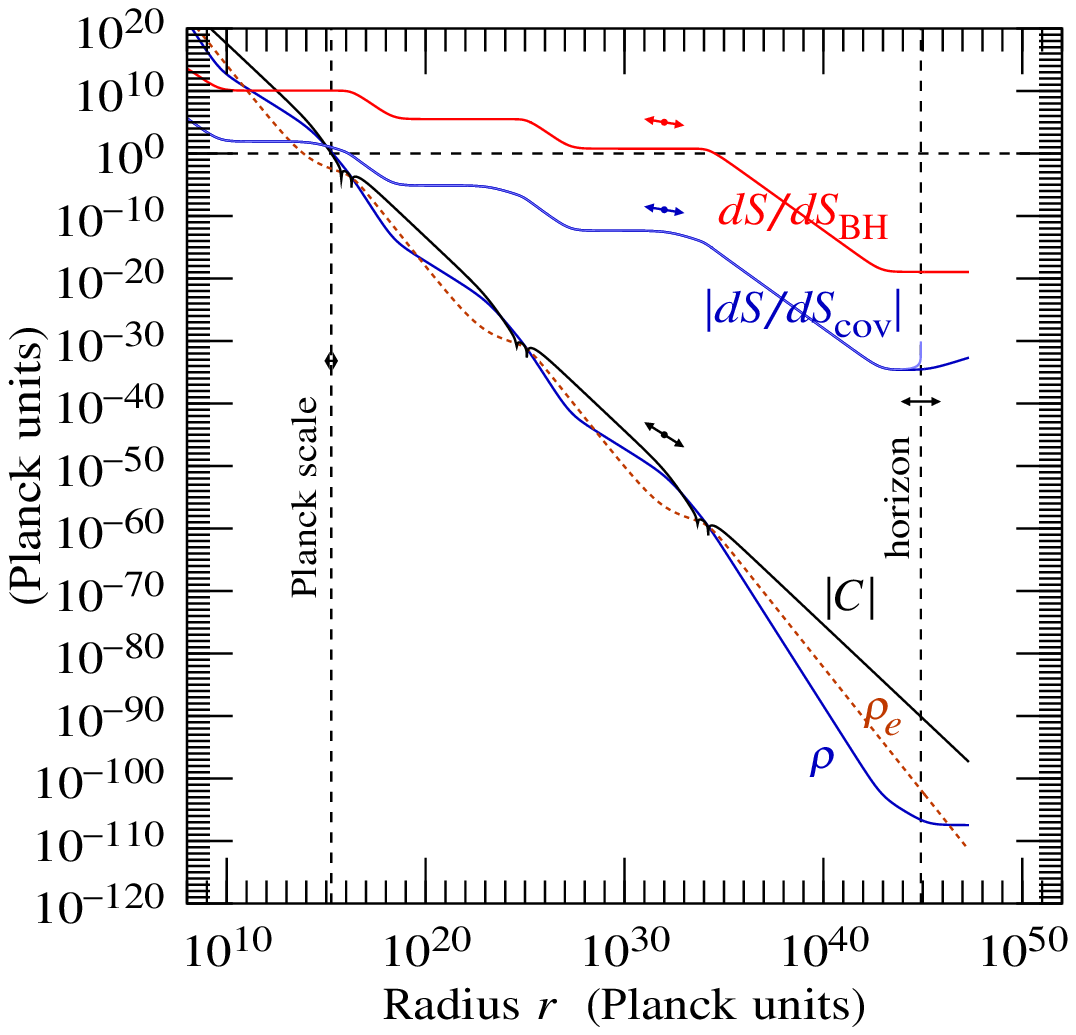}
    \caption[1]{
    \label{vars}
(Color online)
Internal structure of a black hole with
mass $\Mbh = 4 \times 10^6 \unit{\Msun}$,
accretion rate $\Mdot = 10^{-16}$,
charge to mass $\Qbh/\Mbh = 10^{-5}$,
equation of state $w = 0.32$,
and conductivity coefficient $\kappa = 1.24$.
The quantities plotted are,
as a function of radius $r$:
the density $\rho$ of the
baryonic
plasma,
the energy density $\rho_e$ (short dashed line) of the
static
electric field,
the absolute value of
the Weyl curvature scalar
$C = 4 \pi \rho / 3 + Q^2 / ( 2 r^4 ) - M / r^3$,
the rate
$\dd S / \dd \SBH$
of increase
of the black hole entropy with Bekenstein-Hawking entropy,
equation~(\protect\ref{SBH}),
and
the rate
$\lvert \dd S / \dd \Scov \rvert$
at which entropy passes through ingoing (dark)
and outgoing (light) spherical lightsheets
per unit decrease in their area/4,
equation~(\protect\ref{Scov}).
Vertical dashed lines mark the horizon,
and
where the Weyl curvature $\lvert C \rvert$
exceeds $1$ Planck unit.
Arrows,
such as that above
$\dd S / \dd \SBH$,
show how the curves shift
a factor of ten into the past and the future.
The rate
$\dd S / \dd \SBH$
is almost independent of the black hole mass $\Mbh$,
at fixed splat density $\rho_\splat$.
    }
    }
}

\abstract{
We show that the entropy created by Ohmic dissipation inside an  
accreting charged black hole may exceed the Bekenstein-Hawking  
entropy by a large factor.
If the black hole subsequently evaporates, radiating only
the Bekenstein-Hawking entropy,
then the black hole appears to destroy entropy,
violating the second law of thermodynamics.
A companion paper discusses the implications of this startling result.
Bousso's covariant entropy bound is not violated.
}

\keywords{Black Holes, Black Holes in String Theory}

\begin{document}




\section{Introduction}

The purpose of this paper is to show
that classical processes of dissipation can
generate huge quantities of entropy
inside the horizon of a black hole,
many orders of magnitude more than
the Bekenstein-Hawking \cite{Bekenstein:1973ur} entropy.
The specific black hole model presented is intended to be semi-realistic,
albeit over-simplified,
with parameters appropriate to a real supermassive black hole.
We take charge as a surrogate for angular momentum,
and electrical conductivity as a surrogate for angular momentum transport.
To see how much entropy might be created,
we treat the electrical conductivity as an adjustable free parameter.

If a black hole creates many times the Bekenstein-Hawking entropy
and subsequently evaporates,
radiating only the Bekenstein-Hawking entropy
and leaving no remnant,
then entropy is destroyed, violating the second law of thermodynamics.
The implications of this startling result are discussed in a
companion paper \cite{Polhemus:2009}.

Throughout this paper we treat entropy in a purely classical fashion.
In particular, we assume that locality holds inside the black hole.
Locality, the quantum field theory proposition that operators commute
at spacelike-separated points,
is the assumption that normally makes it legitimate to add entropy
over spacelike surfaces.
Since the spacetime curvature inside a supermassive black hole is
well below Planck, except near the singularity,
one might expect classical physics to apply.

It is widely thought that
in order to preserve unitarity of black hole evaporation,
locality must break down
over spacelike surfaces connecting the inside and outside of a black hole
\cite{Susskind:1993if}.
In the companion paper \cite{Polhemus:2009}
we argue that the calculations of the present paper
point to a profligate breakdown of locality inside black holes.

We work in Planck units, $k_B = c = G = \hbar = 1$.

\section{Model of entropy production inside a black hole}

Real supermassive black holes
acquire most of their mass
not during a single collapse event,
but rather by gradual accretion of gas over millions or billions of years.
We 
model this gradual growth 
by the general relativistic,
self-similar, accreting, spherical, charged black hole
solutions described by
\cite{Hamilton:2004av,Hamilton:2004aw},
to which the reader is referred for more detail.
In these models the black hole accretes a charged
``baryonic'' plasma of relativistic matter,
with constant
proper pressure-to-density ratio
$p / \rho = w$,
at a constant rate,
so that the mass of the black hole increases linearly with time.

Real supermassive black holes probably rotate,
but have tiny electric charge.
However, the interior structure of a spherical charged black hole
resembles that of a rotating black hole
in that the negative pressure (tension) of a radial electric field
produces an effective gravitational repulsion
analogous to the centrifugal repulsion inside a rotating black hole.
Thus we 
follow the common practice of taking
charge as a surrogate for spin.
In the self-similar solutions,
the charge of the black hole is produced self-consistently
by the accumulation of the charge of the accreted
plasma.


%

Similarly, we take electrical conduction as a surrogate for the
dissipative transport of angular momentum.
We do not use a realistic electrical conductivity,
but rather treat it as a phenomenological adjustable quantity.
In diffusive electrical conduction,
the electric field $E = Q / r^2$
gives rise to an 
electric current
$j = \sigma E$.
If the conductivity $\sigma$ is taken to be a function only
of the plasma density $\rho$,
then the condition of self-similarity forces
\cite{Hamilton:2004av}
\begin{equation}
\label{sigma}
  \sigma
  =
  {\kappa \rho^{1/2} / ( 4\pi )^{1/2}}
\end{equation}
where the dimensionless conductivity coefficient $\kappa$
is a phenomenological constant.
As discussed by \cite{Hamilton:2004av},
this phenomenological conductivity
is greatly suppressed compared to any realistic conductivity
(except perhaps at densities approaching the Planck density).
However,
angular momentum transport is intrinsically a much weaker process
than electrical conduction,
so it is not unreasonable
to consider a greatly suppressed conductivity.

Since information can propagate only inwards inside a black hole
(at least classically),
it is natural to impose boundary conditions outside the black hole.
We assume that
the
boundary conditions of the accreting black hole
are established at a sonic point outside the horizon,
where the infalling
plasma accelerates
from subsonic to supersonic.
We assume that
the acceleration through the sonic point is finite and differentiable,
which imposes two boundary conditions.
The accretion in real
black holes is likely to be much
more complicated,
but this assumption
is the simplest physically reasonable one.

We define
the charge $\Qbh$
and
mass $\Mbh$
of the black hole at any instant
to be
those
that would be
measured by a distant observer
if there were no charge or mass outside the sonic point,
\begin{equation}
  \Qbh
  =
  Q
  \mbox{ and }
  \Mbh
  =
  M + {Q^2 \over 2 r}
  \mbox{ at the sonic point}
\end{equation}
where
$r$ is the circumferential radius,
and
$Q$ and $M$ denote the interior charge and
mass,
all gauge-invariant scalar quantities.
If the black hole
ceases accreting abruptly at some time,
then $\Qbh$ and $\Mbh$
are the actual charge and mass of the black hole at that time.

Given the assumption that the sonic point is regular,
the dimensionless free parameters of the solutions are:
(1) the mass accretion rate $\Mdot$;
(2) the charge-to-mass ratio $\Qbh / \Mbh$ of the black hole;
(3) the equation of state parameter $w$;
and
(4) the conductivity coefficient $\kappa$.

The black hole mass increases linearly with time,
$\Mbh \propto t$,
and the mass accretion rate $\Mdot$ is
\begin{equation}
\label{Mdot}
  \Mdot
  \equiv
  {\dd \Mbh / \dd t}
  =
  {\Mbh / t}
  \ ,
\end{equation}
where $t = \tau_d = ( r \xi_d^t )_{r = r_s}$
is the time measured by clocks
attached to neutral
dust ($d$)
that free-falls radially
through the sonic point $r = r_s$
from zero velocity at infinity,
and which therefore records the proper time at rest at infinity,
and $\xi_d^t$ is the time component of the homothetic vector
$\xi^k$
in the dust frame
\cite[Appendix~E]{Hamilton:2008zz}.

The density $\rho$ and temperature $T$ of a relativistic fluid
in thermodynamic equilibrium are related by
$\rho = (\pi^2 \geff / 30) T^4$,
where $\geff = g_B + \frac{7}{8} g_F$
is the effective number of relativistic particle species,
with $g_B$ and $g_F$ being the number of bosonic and fermionic species.
If the expected increase in $g$ with temperature $T$ is modeled
(so as not to spoil self-similarity)
as a weak power law
$\geff / g_\Planck = T^\varepsilon$,
with $g_\Planck$ the effective number of relativistic species
at the Planck temperature,
then the relation between density $\rho$ and temperature $T$ is
\begin{equation}
\label{rhoT}
  \rho
  =
  ( {\pi^2 g_\Planck / 30} )
  T^{( 1 + w ) / w}
  \ ,
\end{equation}
with equation of state parameter
$w = 1 / ( 3 + \varepsilon )$
slightly less than
the standard relativistic value $w = 1/3$.
We fix $g_p$ by setting
the number of relativistic particles species to
$g = 5.5$
at $T = 10 \unit{MeV}$,
corresponding to a plasma of relativistic photons, electrons, and positrons.

The entropy $S$ of a proper Lagrangian volume element $\Vol$
of an ideal relativistic fluid
with zero chemical potential is
$S = [ ( \rho \,{+}\, p ) / T ] \Vol$.
The proper velocity of the baryonic fluid
through the sonic point equals the ratio
$\xi^r / \xi^t$
of the radial and time components of the homothetic vector in the
plama
frame
\cite{Hamilton:2004av}.
Thus the entropy $S$
accreted through the sonic point
per unit proper time of the
fluid
is
$\dd S / \dd \tau = [ ( 1 \,{+}\, w ) \rho / T ] 4 \pi r^2 (\xi^r / \xi^t)$.
The sonic radius $r_s$ of the black hole increases
as
$\dd \ln r_s / \dd \tau = 1 / ( r \xi^t )_{r = r_s}$
\cite{Hamilton:2004av}.
The Bekenstein-Hawking entropy of the black hole
is
$\SBH = \pi r_h^2$
where $r_h$ is the horizon radius,
so
$\dd \SBH / \dd \ln r_s = 2 \pi r_h^2$.
Thus the entropy $S$ accreted
per unit increase of the Bekenstein-Hawking entropy is
\begin{equation}
\label{dSaccrete}
  {\dd S \over \dd \SBH}
  =
  {1 \over 2 \pi r_h^2}
  \left.
  {( 1 + w ) \rho 4 \pi r^3 \xi^r \over T}
  \right\rvert_{r = r_s}
  \ .
\end{equation}

Inside the sonic point,
dissipation increases the entropy.
The energy-momentum tensor is the sum of
baryonic
and electromagnetic
parts, $T_b^{mn}$ and $T_e^{mn}$,
and the evolution of baryon entropy is determined by
the time component of
the equation of covariant conservation of energy-momentum
in the rest frame of the baryons:
\begin{equation}
\label{DT}
  D_m T_b^{mt}
  =
  -
  D_m T_e^{mt}
  \ .
\end{equation}
In the self-similar model being considered,
the energy conservation equation~(\ref{DT}) can be written
\cite{Hamilton:2004av}
\begin{equation}
\label{drho}
  {\dd \rho \over \dd \tau}
  +
  ( 1 + w ) \rho
  {\dd \ln ( r^3 \xi^r ) \over \dd \tau}
  =
  \sigma {Q^2 \over r^4}
\end{equation}
which
can be recognized as
an expression of
the first law of thermodynamics
$\dd \rho \Vol + p \dd \Vol = T \dd S$
with proper volume $\Vol \propto r^3 \xi^r$.
The right hand side of equation~(\ref{drho})
is the Ohmic dissipation
$j E = \sigma E^2$.
Equation~(\ref{drho}) can be re-expressed as
\begin{equation}
\label{dlnS}
  {\dd \ln S \over \dd \tau}
  =
  {\sigma Q^2 \over (1 + w) \rho r^4}
\end{equation}
with $
  S
  \propto
  \rho^{1/(1+w)} r^3 \xi^r
  \propto
  ( \rho / T ) r^3 \xi^r
  $.

Since other physics presumably takes over
near the Planck scale,
we truncate the production of entropy at some
arbitrary maximum density $\rho_\splat$
(``rho splat'').
Integrating equation~(\ref{dlnS})
from the sonic point to the splat point
yields the ratio of the entropies
at the sonic and splat points.
Multiplying the accreted entropy, equation~(\ref{dSaccrete}),
by this ratio
yields the rate of increase of the entropy of the black hole,
truncated at the splat point,
per unit increase of its Bekenstein-Hawking entropy
\begin{equation}
\label{SBH}
  {\dd S \over \dd \SBH}
  =
  {1 \over 2 \pi r_h^2}
  \left.
  {( 1 + w ) \rho 4 \pi r^3 \xi^r \over T}
  \right\rvert_{\rho = \rho_\splat}
  \ .
\end{equation}

The 
entropy created becomes large 
when the conductivity coefficient lies
in the range $\kappa \approx 1.3$ to $1000$.
Over this
range
the rate
$\dd S / \dd \SBH$
of increase of entropy, equation~(\ref{SBH}),
is almost independent of the black hole mass $\Mbh$,
\begin{equation}
\label{Sfit}
  {\dd S \over \dd \SBH}
  \approx
  \textrm{const}
  \approx
  {
  2
  ( 1 - \Qbh^2 / \Mbh^2 )^{1/2}
  ( 1 + w )
  \rho_\splat
  \over
  \Mdot
  \bigl[ 1 + ( 1 - \Qbh^2/\Mbh^2 )^{1/2} \bigr]^2
  \sigma_\splat
  T_\splat}
\end{equation}
in which the empirical fit on the right hand side
is accurate to
a factor of two
over the range
$\kappa \approx 10$ to $1000$
(for $\kappa \lesssim 10$ to $\approx 1.3$,
the fit increasingly overestimates
$\dd S / \dd \SBH$),
and
$\Mbh \gtrsim 3 \unit{\Msun}$,
$\Mdot \lesssim 10^{-4}$,
$\Qbh / \Mbh \approx 10^{-12}$ to $0.99999$,
$w \approx 0.1$ to $0.55$,
and $\rho_\splat$ not too small.

Bousso
\cite{Bousso:2002ju}
has proposed the covariant entropy bound,
which states that the entropy
passed through a converging lightsheet
cannot exceed
$\Scov \equiv \textrm{area}/4$
of its boundary.
In the models under consideration,
the rate at which entropy passes through an ingoing or outgoing spherical lightsheet
per unit decrease in
$\Scov = \pi r^2$ is
\begin{equation}
\label{Scov}
  \left\lvert {\dd S \over \dd \Scov} \right\rvert
  =
  {\dd S \over \dd \SBH}
  {r_h^2 \over r^2}
  {1 \over \xi^r \lvert \beta \mp \gamma \rvert}
\end{equation}
in which
$\{ \beta, \gamma \}$
are the time and radial components
of the proper covariant radial 4-gradient in the notation of \cite{Hamilton:2004av},
and
the $\mp$ sign is $-$ for ingoing, $+$ for outgoing
lightsheets.
A sufficient condition for the covariant entropy bound to be satisfied is
$\lvert \dd S / \dd \Scov \rvert \leq 1$.

\varsfig

\section{Example of interior entropy exceeding Bekenstein-Hawking}
\label{example}

A black hole of mass
$4 \times 10^6 \unit{\Msun} \approx 4 \times 10^{44}$ Planck units
(the mass of the supermassive black hole at the center of the Milky Way
\cite{Ghez:2003qj,
Eisenhauer:2005cv})
accreting over the age of the Universe
$10^{10} \unit{yr} \approx 6 \times 10^{60}$ Planck units
has an accretion rate of $\Mdot \approx 10^{-16}$.
Figure~\ref{vars}
shows the interior structure of a black hole
with that mass and accretion rate,
charge-to-mass $\Qbh / \Mbh = 10^{-5}$,
equation of state $w = 0.32$,
and conductivity coefficient $\kappa = 1.24$.

To produce lots of entropy,
the baryonic plasma must fall to a central singularity,
and we choose the conductivity
$\kappa = 1.24$
to be
at (within numerical accuracy)
the critical conductivity
\cite{Hamilton:2004av}
for this to occur.
Below the critical conductivity
the plasma generally does not fall to a singularity,
but rather drops through the Cauchy horizon.
The latter solutions are subject
to the mass inflation instability
\cite{Poisson:1990eh,Hamilton:2004aw},
a fascinating regime not considered in this paper.

Solutions at the critical conductivity exhibit
\cite{Hamilton:2004av}
the periodic self-similar behavior
first discovered by \cite{Choptuik:1992jv}.
The ringing of the curves shown in Figure~\ref{vars}
is a manifestation of this,
not a numerical error.

The electric charge advected by the plasma inwards across the sonic point
is, thanks to the ``high'' conductivity,
almost canceled by an outward current.
As a consequence,
the charge-to-energy of the accreted plasma,
here
$\approx 400$
at the sonic point,
is substantially
larger than the charge-to-mass of the black hole.
We have deliberately chosen a small charge-to-mass ratio for the black hole,
$\Qbh / \Mbh = 10^{-5}$,
so that the Lorentz repulsion of the plasma by the black hole is subdominant,
and the trajectories of parcels of plasma
outside the black hole are not greatly different from Schwarzschild geodesics.
Thus for example
the sonic point is at a radius of $3.06$ geometric units
($c = G = \Mbh = 1$),
close to that expected for a neutral tracer relativistic fluid that
free-falls from zero velocity at infinity.
The horizon is at $2.00$ geometric units,
like Schwarzschild.
Figure~\ref{vars}
shows the solution out to $2{,}000$ geometric units.

At the sonic point,
the plasma temperature is
$\approx 4 \,{\times}\, 10^5 \unit{K}$.
Inside the horizon,
the electric field increases,
and Ohmic dissipation starts to heat the plasma,
increasing its temperature and entropy.
When the plasma energy has become comparable to the electric energy,
then the plasma goes into a power law regime
where the plasma and electric energies increase in proportion to each other,
kept in lockstep by the conductivity.

The entropy hits the Bekenstein-Hawking milestone,
$\dd S / \dd \SBH = 1$,
when
the temperature is $\approx 3 \times 10^{-16}$ Planck units,
or $3 \unit{TeV}$,
and
the curvature radius is
$\lvert C \rvert^{-1/2} \approx 10^{30}$ Planck lengths,
or $0.01 \unit{mm}$.
This temperature and curvature are
almost independent of the mass $\Mbh$
of the black hole,
equation~(\ref{Sfit}).

If the plasma's dissipative trajectory is followed to the Planck scale,
$\lvert C \rvert = 1$,
then the rate of increase of entropy
relative to Bekenstein-Hawking is
$\dd S / \dd \SBH \approx 10^{10}$,
again
almost independent of the mass $\Mbh$ of the black hole.
If the entropy is assumed to accumulate additively inside the black hole
then the cumulative entropy can evidently
exceed Bekenstein-Hawking by a large factor.

Figure~\ref{vars}
shows that
$\lvert \dd S / \dd \Scov \rvert \leq 1$,
equation~(\ref{Scov}), at all sub-Planck scales.
Thus although the cumulative entropy may exceed Bekenstein-Hawking,
Bousso's covariant entropy bound is satisfied by the black hole.

\section{Conclusion}
\label{summary}

We have shown that
the dissipation of the free energy of the electric field
inside a charged black hole
can potentially create many
times more than the Bekenstein-Hawking entropy.
If the black hole subsequently evaporates,
radiating only the Bekenstein-Hawking entropy
and leaving no remnant,
then
entropy is destroyed.

This startling conclusion is premised on the assumption
that entropy created inside a black hole
accumulates additively on spacelike slices,
which in turn derives from the assumption that the Hilbert space
of states is multiplicative over spacelike-separated regions,
as postulated by locality.
This is essentially the same reasoning
that originally led Hawking
\cite{Hawking:1976ra}
to conclude
that black hole evaporation is non-unitary,
that black holes must destroy information.

It is widely thought that unitarity should be considered a
higher principle than locality.
To ensure that black hole evaporation is unitary,
locality between the inside and outside of a black hole
must break down.
In a companion paper \cite{Polhemus:2009}
we argue that the gross violation of the second law
found in the present paper
points to a wholesale breakdown of locality inside black holes,
and provides a compelling argument in favor of the conjecture
of ``observer complementarity''.

The black hole respects Bousso's
covariant entropy bound
\cite{Bousso:2002ju},
as it should given the theorem of
\cite{Flanagan:1999jp}.

\section*{Acknowledgements}
This work was supported by
NSF award
AST-0708607.
We thank the University of Colorado's High Energy Physics Group,
without whom this paper would not exist.

\bibliographystyle{JHEP}
\bibliography{BlackEigen}

\end{document}